 \documentclass[11pt, a4paper]{article}
\usepackage{geometry}
\usepackage[english]{babel}
\usepackage{pdflscape}

\usepackage[round, sort]{natbib}
\usepackage{booktabs}

\usepackage{float}

\usepackage{caption}
\usepackage{threeparttable}

\usepackage{pdflscape}

\usepackage{siunitx}
\newcommand{\sym}[1]{\rlap{\textsuperscript{#1}}} % for significance stars without affecting alignment

\usepackage{amsmath}
\usepackage{graphicx}
\usepackage{subcaption}
\usepackage[colorlinks=true, allcolors=blue]{hyperref}
\usepackage{orcidlink} % for ORCID links
\usepackage{makecell}
\usepackage{tabularx}

\usepackage{multirow}

\usepackage{multicol}

\usepackage{rotating}

\newcommand{\keywords}[1]{%
\begin{center}
\begin{minipage}{0.92\textwidth}\small
\textbf{Keywords.} #1
\end{minipage}
\end{center}
}

\newif\ifanonymous
\anonymousfalse % \anonymoustrue \anonymousfalse

\title{\textit{Preprint:}\\Would I regret being different? The influence of social norms on attitudes toward AI usage}
\ifanonymous
    \author{Anonymous Author(s)}
\else
    \author{
        Jaroslaw Kornowicz \orcidlink{0000-0002-5654-9911}\thanks{Corresponding author.}\\
        Paderborn University, Paderborn, Germany\\
        \href{mailto:jaroslaw.kornowicz@uni-paderborn.de}{jaroslaw.kornowicz@uni-paderborn.de}
          \and
        Maurice Pape \orcidlink{0009-0001-5627-8808}\\
        Paderborn University, Paderborn, Germany\\
        \href{mailto:iammlp@campus.uni-paderborn.de} {iammlp@campus.uni-paderborn.de}
        \and
        Kirsten Thommes \orcidlink{0000-0002-8057-7162}\\
        Paderborn University, Paderborn, Germany\\
        \href{mailto:kirsten.thommes@uni-paderborn.de}{kirsten.thommes@uni-paderborn.de}
    }
\fi

\begin{document}

\maketitle

\begin{abstract}
% Please provide an abstract of 150 to 250 words. The abstract should not contain any undefined abbreviations or unspecified references.
Prior research shows that social norms can reduce algorithm aversion, but little is known about how such norms become established. Most accounts emphasize technological and individual determinants, yet AI adoption unfolds within organizational social contexts shaped by peers and supervisors. We ask whether the source of the norm---peers or supervisors---shapes AI usage behavior. This question is practically relevant for organizations seeking to promote effective AI adoption. We conducted an online vignette experiment, complemented by qualitative data on participants’ feelings and justifications after (counter-)normative behavior. In line with the theory, counter-normative choices elicited higher regret than norm-adherent choices. On average, choosing AI increased regret compared to choosing an human. This aversion was weaker when AI use was presented as the prevailing norm, indicating a statistically significant interaction between AI use and an AI-favoring norm. Participants also attributed less blame to technology than to humans, which increased regret when AI was chosen over human expertise. Both peer and supervisor influence emerged as relevant factors, though contrary to expectations they did not significantly affect regret. Our findings suggest that regret aversion, embedded in social norms, is a central mechanism driving imitation in AI-related decision-making.

\end{abstract}

\keywords{Social norms, Algorithm aversion, Decision support systems, Human-AI interaction, Vignette experiment}

% \statementsanddeclarations

% \vspace{5cm}
% \ifanonymous\else
% \begin{center}
% \begin{minipage}{0.92\textwidth}\small
% \textbf{Funding.} The authors gratefully acknowledges funding by the German Research Foundation (Deutsche
% Forschungsgemeinschaft, DFG): TRR 318/1 2021 – 438445824
% \end{minipage}
% \end{center}
% \fi

% \begin{center}
%     \textit{A preprint version of the paper is hosted on XXX}
% \end{center}

\newpage

% \begin{multicols}{2}

\section{Introduction}
Consider the case of a newly hired employee at a management consulting firm who is supposed to conduct a market analysis. The employee faces a choice between consulting a senior analyst or relying on an AI tool such as ChatGPT. This situation reflects a broader research question: under what conditions do individuals rely on AI, and when do they prefer human expertise?

Prior work has identified two opposing patterns. Algorithm aversion describes reluctance to trust or rely on AI or algorithmic advisors \citep{Dietvorst_Simmons_Massey_2015, kornowicz2025algorithm}, whereas algorithm appreciation refers to contexts in which individuals prefer AI over human judgment \citep{Logg_Minson_Moore_2019}. Although AI can outperform humans in many domains, its guidance is frequently met with skepticism or rejection. This tension has led scholars to examine strategies for reducing algorithm aversion and improving decision-making quality \citep{Mahmud_Islam_Mitra_2023}.

Determinants of AI reliance can be grouped into two categories. The first concerns technological characteristics, such as performance and reliability \citep{Jussupow}. The second concerns user characteristics, such as gender or personality traits \citep{kaplan2023trust, Jussupow}. Interventions, therefore, target both dimensions, ranging from improving AI transparency \citep{Poursabzi_2021} to tailoring systems to user preferences \citep{Lai_Zhang_Chen_Liao_Tan_2023}.

These perspectives, however, implicitly assume that reliance is driven solely by properties of the technology or the individual. In practice, AI adoption occurs within a broader social context. Decisions are shaped by organizational environments, social expectations, and the presence of observers. For example, a new employee’s task performance is evaluated not only by their own judgment but also by supervisors and peers.

Recent research has highlighted the role of social norms in AI adoption. Social norms—defined as perceptions of what is typical or desirable within a group—are powerful determinants of behavior \citep{Cialdini_Goldstein_2004}. They can be distinguished as descriptive norms (what others do) and injunctive norms (what others approve of) \citep{Cialdini_2003}. Evidence suggests that descriptive norms exert a particularly strong influence. For example, \cite{alexander2018trust} found that observing others use AI increased adoption more than learning about its accuracy. \cite{Bogard_Shu_2022} also showed that algorithm aversion could be reversed when the social norm favored algorithm usage, implying that adoption accelerates once a critical mass of peers has begun using AI, highlighting the importance of social proof.

In organizational settings, employees do not simply observe others; they also interpret normative information selectively \citep{Tankard_Paluck_2016}. Norm influence varies depending on the source. Individuals are more strongly influenced by salient peers \citep{paluck2012salience} and group leaders \citep{Hogg_2010, Robertson_Barling_2013}. Interestingly, new employees sometimes conform more to lower-ranking colleagues than to higher-ranking ones, perceiving the former as more similar and thus more relevant as reference points \citep{Dannals_Reit_Miller_2020}. Similarity is therefore a central factor in norm influence \citep{Bicchieri_Dimant_2022}. In general, norms from close and relatable sources (e.g., friends, family) are more influential than those from distant authorities (e.g., doctors, priests, officials) \citep{Melnyk_van}.

At the same time, organizational evaluations are often made by leaders rather than peers. Under uncertainty about which norms apply and the consequences of violating them, following leaders’ behavior may be more rational.

This leads to the present research question. While \cite{Bogard_Shu_2022} demonstrated that social norms can reduce algorithm aversion, they did not explore how algorithms come to be accepted as the norm. Therefore, we specifically ask: \textit{What role does the source of the norm play in shaping AI usage behavior?}

This question has practical implications. Should organizations rely on early adopters to establish descriptive norms through peer influence, or should leaders model AI adoption themselves to set organizational standards? Because higher levels of AI adoption can improve decision quality, understanding how norms are best established is of direct relevance to organizational practice.

To address this question, we conducted an online behavioral experiment. Building on the paradigm developed by \cite{Bogard_Shu_2022}, participants were presented with randomized vignettes in which they assumed the role of a newly hired analyst. The quantitative part of our experiment was complemented by a qualitative part that explored participants’ feelings and justifications following (counter-)normative behavior.

Following Regret Theory \citep{Bell1982, loomes1982regret}, we define \textit{regret} as the counterfactual negative emotion felt upon learning that a foregone option would have yielded a better outcome. Our results show that counter-normative behavior is more likely to elicit regret than normative behavior, helping to explain the reluctance of early adopters. Choosing AI, in particular, amplified regret; however, this aversion weakened when AI use was framed as the prevailing norm, indicating an interaction between choosing AI and an AI-favoring norm. Moreover, participants tended to assign less blame to technology than to humans, which in turn heightened regret when AI was chosen over human expertise. Finally, both peer and supervisor influence mattered for technology adoption; contrary to expectations, their effects on regret were comparable. Taken together, our findings suggest that regret aversion, shaped by social norms, is a central driver of imitation in decision-making. In the following section, we theorize how social norms influence technology use and present the methods and results of our study.

\section{Social Norms and Technology Adoption}

Social norms play a central role in shaping both the adoption and use of new technologies \citep{venkatesh2000don, huttel2022importance}. At the same time, technologies can reshape social norms. For instance, surveillance practices that were once considered intrusive or ``creepy'' have become normalized in the context of childcare, home monitoring, and pet tracking \citep{tene2013theory}. Ultimately, individuals’ perceptions of technology-related social norms strongly influence their willingness to adopt or reject innovations.

Two types of social norms are particularly relevant. Descriptive norms emerge when individuals observe the behavior of others in a given situation \citep{koonce2015effects}. They are grounded in informational social influence and are generally straightforward to interpret \citep{mcdonald2015social}. In contrast, injunctive norms arise from perceptions of what behaviors others approve or disapprove of \citep{Cialdini_Kallgren_Reno_1991}. Injunctive norms are more complex because they require individuals to anticipate peers’ evaluations. For example, in an online meeting, if everyone activates their camera, the descriptive norm is to keep the camera on. The injunctive norm, however, depends on how peers might judge someone who decides to turn it off.

Normative influence is therefore closely tied to peers but can generate conflict when personal or moral norms diverge from group expectations \citep{Basic_Verrina_2024}. When personal and group norms align, adherence is reinforced. When they conflict, individuals face a trade-off between violating their own beliefs and bearing the psychological or social costs of deviating from group norms \citep{michaeli2015norm}. Norm violations often trigger regret \citep{loomes1982regret}, and individuals sometimes forgo monetary benefits to avoid this negative emotion \citep{Bell1982}.

Regret can arise both from internal disappointment and from anticipated sanctions. Norm violations expose individuals to social costs, ranging from mild disapproval to outright exclusion \citep{fehr2018social}. Such sanctions can be imposed by peers, third parties, or institutions \citep{Gelfand_Gavrilets_Nunn_2024}. Even weak punishment can reinforce norms if people are aware of the behavioral standard \citep{bicchieri2021deviant}. 

Norms are particularly influential under conditions of uncertainty, when appropriate behavior is ambiguous. In such contexts, individuals rely more heavily on social cues and feel stronger emotional reactions when outcomes deviate from what is perceived as normal \citep{Kahneman_Miller_1986, Feldman_Albarracín_2017, roese1997counterfactual}. Newcomers to a group or workplace, for example, must learn and internalize new norms before acting with confidence \citep{zhang2023we}. During this adjustment, new tools such as algorithms may initially appear to be norm violations, further complicating adoption \citep{Bogard_Shu_2022}.

Technology adoption is therefore not merely a function of technical performance but also of social context. Peers often adopt similar technologies \citep{dickinger2008role, osatuyi2019social}, and social influence is particularly salient in the diffusion of AI \citep{strzelecki2024use, zheng2024examining}. Importantly, AI diffusion is not always correlated with quality. Although AI often outperforms humans across domains \citep{cheng2016computer, luo2019frontiers}, individuals frequently display algorithm aversion---a tendency to distrust AI following errors \citep{Dietvorst_Simmons_Massey_2015}. Errors by algorithms are penalized more harshly than comparable human errors, especially when they occur early \citep{prahl2017understanding, gill2024dynamics, kim2023algorithms}. Even when explicitly informed about AI’s superior performance, individuals often remain indifferent between AI and human advice rather than preferring the algorithm \citep{Castelo_Bos_Lehmann_2019}. This may stem from unrealistic expectations of perfection \citep{madhavan2007similarities}, as seen in patients who trusted physicians less when they used computer-aided diagnostics compared to colleagues or independent decision-making \citep{shaffer2013patients, orban2025trust}.

Nonetheless, interventions can mitigate algorithm aversion. Allowing users to modify algorithms \citep{dietvorst2018overcoming} or framing them as ``learning systems'' \citep{berger2021watch, chacon2025preventing} increases acceptance. Moreover, aversion is less likely when tasks are perceived as objective or numerical, sometimes even leading to algorithm appreciation \citep{Logg_Minson_Moore_2019, orban2025trust}. Context and framing, therefore, matter: both aversion and appreciation are possible depending on the technology, the individual, and the situation \citep{hou2021expert, huynh2025generative}. Familiarity is also key—trust in algorithms often increases with repeated exposure \citep{jussupow2024integrative, mahmud2024decoding, ngo2025humanizing}.

A related perspective is the Regret Theory, which predicts stronger adverse reactions when bad outcomes result from counter-normative behavior compared to norm-conforming actions \citep{Kahneman_Miller_1986, Feldman_Albarracín_2017, tuncc2025regret}. Counterfactuals are easier to imagine when decisions deviate from norms, amplifying regret. For example, juries award higher compensation when harm arises from unusual rather than normal circumstances \citep{Miller_McFarland_1986}. Similarly, in technology adoption, regret is likely to be intensified when unfavorable outcomes follow norm violations rather than norm adherence \citep{Bogard_Shu_2022, tuncc2025regret}.

This logic suggests three implications. First, individuals are more likely to adopt technologies that are already widely used by peers \citep{alexander2018trust}, particularly those they perceive as similar to themselves \citep{Dannals_Reit_Miller_2020}. Second, regret-minimization strategies \citep{Simonson_1992, zeelenberg2004consequences} encourage adherence to social norms, even at the expense of monetary gains \citep{Bell1982}. Third, counter-normative adoption of AI is particularly likely to generate regret, whereas norm-aligned adoption can foster appreciation \citep{Bogard_Shu_2022}.

From this reasoning, we derive the following hypotheses:

\begin{description}
\item[H1:] Regret is higher when one makes a counter-normative decision\citep{Feldman_Albarracín_2017, Kahneman_Miller_1986, tuncc2025regret}.
\item[H2:] Regret is higher when one chooses help from an AI tool instead of a human \citep{Dietvorst_Simmons_Massey_2015}.
\item[H3:] Counter-normative behavior has a greater effect on regret than using an AI tool \citep{Bogard_Shu_2022}.
\item[H4:] Regret is lower when one chooses help from an AI tool if the AI tool represents the norm \citep{Bogard_Shu_2022}.
\end{description}

Finally, the source of the norm is also critical. Individuals are more likely to conform when the source is perceived as similar or proximate \citep{Bicchieri_Dimant_2022, Dannals_Reit_Miller_2020}. However, superiors may exert stronger influence than peers for two reasons. First, they can signal both descriptive and injunctive norms; second, they can sanction non-compliance by virtue of their authority \citep{Blanton_Christie_2003, millerprentice2013psychological}. Yet, authority does not always outweigh closeness, as norms from relatable peers sometimes prove more influential \citep{Melnyk_van, Melnyk2022}. Based on this tension, we hypothesize:

\begin{description}
\item[H5:] Regret is higher when the norm source is a superior.
\end{description}

\section{Method}
%\subsection{Overview}
To address our research question on the role of norms concerning the use of AI support, a between-subjects experimental vignette study was conducted online \citep{Aguinis_Bradley_2014}. The vignette study placed participants in a scenario that allowed the examination of norms within a professional context. The participants are placed as newly hired employees at a consulting firm, where they are tasked with conducting a market analysis. We deliberately use a newcomer frame because norm effects are strongest under uncertainty \citep{smith2007uncertainty}. In this scenario, they are uncertain about how to proceed and see two options: either ask a \textit{senior analyst} or use an \textit{AI tool}. During this process, they observe either their \textit{superior} or \textit{colleagues} and, in doing so, perceive which option is considered the norm. A multifactorial treatment design, based on the experiments by \cite{Bogard_Shu_2022}, was employed, resulting in different vignette conditions. Regret serves as the dependent variable, which participants report based on their decision to either follow the norm or not, as presented in the vignette, after learning that their market analysis was not well received by clients. 

% Participants were compensated with a fixed payment of £2 for their involvement.

The section \nameref{sec:method_conditions} explains the design of the vignette. From the participant's perspective, the experiment's procedure is detailed in \nameref{sec:mathod_procedure}. Finally, the recruitment of participants is discussed in \nameref{sec:method_participants}.
The experiment was preregistered before data collection\footnote{\url{https://osf.io/a3jwv/?view_only=4d101fe0db554668b23691f26100a84e}}. The ethics board of the University of Paderborn approved the research project.

\subsection{Experimental Conditions and Vignette}
\label{sec:method_conditions}

A 2x2x2 treatment design was used, with the following factors:

\begin{description} 
    \item[Norm adherence:] Whether the employee’s action conforms to the specified \textbf{Norm}. When the \textbf{Norm} is \textit{senior analyst}, norm-adherent behavior is to consult the \textit{senior analyst}; choosing the \textit{AI tool} is counter-normative. When the \textbf{Norm} is \textit{AI tool}, norm-adherent behavior is to use the \textit{AI tool}; consulting the \textit{senior analyst} is counter-normative.

    \item[Norm:] Which behavior is considered the norm---consulting a \textit{senior analyst} or using an \textit{AI tool}.
    \item[Source of the Norm:] Who is being observed by the employee---either a \textit{superior} or \textit{colleagues}.
\end{description}

The following vignette was presented to the participants, with the text adjusted to match their respective treatment conditions. Additionally, comic illustrations were added to support the vignette. Screenshots of one of the eight conditions can be found in Appendix \ref{appendix:screenshots}.

\textit{``Imagine you are a new employee at a consulting firm. Your first task is to create a market analysis report for a client, as assigned by your superior manager. The report should include an overview of industry trends, a review of potential growth opportunities, and a set of actionable recommendations. Although you have some background knowledge, you feel unsure about the most effective way to tackle the project. You know you have two options: you can either use an AI tool (such as ChatGPT, Copilot, or Gemini) or seek guidance from a senior analyst. As you are not sure what to do, you walk around to see how others are working. You observe that your superior manager [colleagues] usually ask[s] an AI tool [a senior analyst] for advice on how to approach similar projects. After weighing your options, you decide to ask a senior analyst instead of using an AI tool [use an AI tool instead of asking a senior analyst] for help with this assignment. The senior analyst [AI tool] provides guidance and input, and you finalize the report using the information provided by the analyst [AI tool]. Feeling more confident after following this approach, you submit the report to the client. However, the client later provides negative feedback. They mentioned that the report lacked critical insights and did not meet the expected quality standards. They express disappointment and note that it did not fully address key areas they were interested in.''}

% How much do you expect to regret choosing to use the (Decision: senior analyst / AI tool) for this task? (1 = Not at all, 7 = Very much) How likely would you be to use the (Decision: senior analyst / AI tool) for future projects? (1 = Very unlikely, 7 = Very likely)''

% \subsection{Dependent Variable}
% \label{sec:method_hypotheses}

% We follow the approach of \cite{Bogard_Shu_2022} and use expected \textit{regret} as the dependent variable to examine the effects of norms and their origins. This variable is measured using a 7-point Likert scale after participants have read the vignette. 

\subsection{Procedure}
\label{sec:mathod_procedure}

The experiment was conducted in 2025 online using oTree software \citep{chen2016otree}. Participants were recruited via Prolific.com and redirected to the study platform.

\subsubsection{Study Introduction}
Upon entry, participants provided their Prolific ID, read the privacy policy, and were randomly assigned to one of eight experimental conditions. The study began with a general instruction screen (see Appendix \ref{appendix:instructions}), followed by four comprehension questions. Each question allowed a maximum of two incorrect attempts. Participants who failed at least one question three times were disqualified, instructed to return their submission on Prolific, and excluded from subsequent analyses.

\subsubsection{Vignette and Regret Measurement}
Participants who passed the comprehension stage were presented with a vignette that varied according to the assigned condition. After reading the vignette, participants reported the regret they experienced in the role of the employee. Regret was measured using the validated Decision Regret Scale \citep{Brehaut_2003}, which consists of five items. This measure extends prior work by \cite{Bogard_Shu_2022}, who relied on a single-item approach.

\subsubsection{Open-Ended Questions}
Following the regret measure, participants answered two open-ended questions. First, they described the factors that influenced their decision in the vignette. Second, they explained how they would justify this decision to their manager. Responses were collected in mandatory open-text fields without character limits, enabling richer insights beyond the standardized scale. In addition, an inductive coding of the collected texts was conducted. Themes were developed independently for both questions in an inductive process, resulting in different themes for the questions. Each text was assigned to one or more themes in an iterative process. The common themes were then combined into broader groups to provide a better overview of the most important aspects. An example of how the answers were characterized is given in Table \ref{tab:labels}, the full coding scheme with examples can be found in Appendix \ref{appendix:Coding}. 

\begin{table}[h!]
\centering
\caption{Example for the assorting of themes for the open-ended questions.}
\begin{tabular}{p{0.65\textwidth} p{0.3\textwidth}}
\toprule
\textbf{Answers} & \textbf{Themes} \\
\midrule
I regret using the AI tool as my manager was disappointed with my final result and if I had asked a senior analyst instead, I may have received better and more personalised guidance with someone who understands my exact job needs &  Better output by expert, Regrets decision \\
Going against what was standard company procedure - team members asking an analyst. The negative response by the customer The lack of detail mentioned in the feedback &  Counter-normative decision, Disappointed client\\
i was new i thought someone that as worked here a long time might be able to help me better & Insecure, Overestimate analyst, Lack of competence \\
I used AI as I thought it could do the job of a senior analyst. I was wrong. & Overestimate AI, Responsibility orientation \\
\bottomrule
\label{tab:labels}
\end{tabular}
\end{table}

\subsubsection{Attention Check \& Demographic Data}
Finally, participants completed an attention check. Without access to the vignette, they answered three multiple-choice questions about (1) who established the norm in the scenario, (2) what the norm entailed, and (3) the decision they had made. Only participants who answered all three correctly were retained for analysis. In the last part of the study, participants completed a brief demographic questionnaire before being redirected to Prolific.

\subsection{Participants}
\label{sec:method_participants}
%The study was conducted in 2025. Participants were recruited from the platform Prolific.com.
\subsection{Sample Size Determination and Participants}

Prior to data collection, the required sample size was determined using a power analysis in G*Power \citep{gpower}. To adjust for multiple hypothesis testing, a Holm-Bonferroni correction was applied \citep{Holm_1979}. We specified a medium effect size of $f = 0.25$, a significance level of $\alpha = 0.01$, and a statistical power of $1 - \beta = 0.90$. The corrected alpha level reflects the five planned hypotheses. With eight experimental groups, the analysis indicated a minimum required sample size of 242 participants.

Eligibility criteria for participants from prolific were an approval rate above 99\%, completion of at least 10 prior studies, fluency in English, and current residence in the United Kingdom. These restrictions were imposed to ensure comprehension of the vignette and the validity of responses. Participants received a fixed payment of £2.

A total of 316 individuals completed the study. Of these, 59 participants (18.67\%) failed the attention check and were excluded from all analyses. An additional 12 participants (4.67\%) were excluded for completing the vignette page in an implausibly short time. The final analytic sample comprised 245 participants. The mean age was 43.24 years, and 53.47\% of the sample identified as male. The average completion time was approximately 11 minutes, corresponding to an effective hourly compensation of £13.10.

All experimental data (excluding personally identifying information), along with the analysis code, are provided in the online appendix\footnote{\url{https://osf.io/u6wkg/?view_only=87666c46dc614e71904339968905fc20}}. Analyses were conducted using Python and STATA; a full list of software packages and version numbers is included in the online data repository.

\section{Results}
\subsection{Regret Measure}
The dependent variable in the quantitative analysis was \textit{regret}, measured by participants’ self-reports in the role of the employee described in the vignette. Responses were collected on a 5-point Likert-type scale. A principal-axis exploratory factor analysis constrained to a single factor yielded loadings between 0.56 and 0.88 ($M = 0.82$) after reverse-scoring negatively keyed items. This factor accounted for 63\% of the variance, supporting the assumption of unidimensionality. Internal consistency was excellent (Cronbach’s $\alpha = 0.89$; McDonald’s $\omega = 0.89$). All corrected item–total correlations exceeded 0.50, and removing any single item altered $\alpha$ by no more than 0.02. Accordingly, all five items were retained, and their mean was used to compute the composite regret score. Table \ref{tab:regret_treatments} presents the distribution of participants across the eight treatment conditions, together with the corresponding average regret scores and standard deviations.

\begin{table}[ht]
\centering
\caption{\textit{Regret} scores by treatment.}
\label{tab:regret_treatments}
\begin{tabular}{ccccc}
\toprule
\textbf{Norm Adherence} & \textbf{Norm} & \textbf{Source} & \textbf{$n$} & \textbf{Mean \& SD Regret} \\
\midrule
No   & Analyst & Colleagues & 33 & $4.33 \pm 0.45$ \\
No   & Analyst & Superior   & 22 & $4.30 \pm 0.46$ \\
Yes  & AI      & Superior   & 29 & $4.01 \pm 0.85$ \\
Yes  & AI      & Colleagues & 36 & $3.82 \pm 0.68$ \\
No   & AI      & Superior   & 32 & $3.42 \pm 0.94$ \\
Yes  & Analyst & Superior   & 24 & $3.28 \pm 0.88$ \\
Yes  & Analyst & Colleagues & 33 & $3.13 \pm 0.85$ \\
No   & AI      & Colleagues & 36 & $3.10 \pm 0.86$ \\
\bottomrule
\end{tabular}
\end{table}

% HABEN WIR AUCH PAARWEISE VERGELICHE? FALLS JA; WÜRDE ICH DAS ERSTMAL HIER ERZÄHLEN.
Appendix \ref{appendix:pairwise} shows the pairwise comparisons of all eight groups using a t-test. To test the five hypotheses, two linear regression models are employed:

\begin{align}
\text{Regret}_{i} &= \beta_{0}
  + \beta_{1}\,\text{CounterNorm}_{i}
  + \beta_{2}\,\text{UsedAI}_{i}
  + \beta_{3}\,\text{SourceSuperior}_{i}
  + \delta Z_{i}
  + \varepsilon_{i} \\[6pt]
\text{Regret}_{i} &= \beta_{0}
  + \beta_{1}\,\text{UsedAI}_{i}
  + \beta_{2}\,\text{NormAI}_{i}
  + \beta_{3}\,(\text{UsedAI}\times\text{NormAI})_{i}
  + \delta Z_{i}
  + \varepsilon_{i}
\end{align}

The independent variable \textit{NormAI} indicates whether using AI is the norm in the vignette. \textit{CounterNorm} captures whether the employee in the vignette deviated from the norm---in other words, it reflects the opposite of norm adherence. Based on this, the variable \textit{UsedAI} denotes whether the employee used AI for the task, and \textit{SourceSuperior} indicates whether the source of the norm is the superior or colleagues. Finally, \textit{Z} represents the set of control variables (sex, age, and university education). We regress the independent and control variables on regret (Table \ref{tab:regression_models}).\\

\begin{table}[!htbp]\centering
\begin{threeparttable}
\caption{\textit{Regret} Regression Models}
\label{tab:regression_models}

\setlength{\tabcolsep}{8pt}
\sisetup{
  table-number-alignment = center,
  table-figures-integer = 1,
  table-figures-decimal = 3,
  table-figures-exponent = 0
}

\begin{tabular}{
  l
  S[table-format=1.3]
  @{\,\phantom{\textsuperscript{***}}(} S[table-format=1.3] @{)}
  @{\hspace{1.0em}}
  S[table-format=1.3]
  @{\,\phantom{\textsuperscript{***}}(} S[table-format=1.3] @{)}
}
\toprule
& \multicolumn{2}{c}{(1)} & \multicolumn{2}{c}{(2)}\\
\midrule
\textit{CounterNorm}                         &  0.252\sym{**}  & 0.099 & \multicolumn{2}{c}{} \\
\textit{UsedAI}                              &  0.887\sym{***} & 0.096 &  1.138\sym{***} & 0.127 \\
\textit{SourceSuperior}                      &  0.162          & 0.100 & \multicolumn{2}{c}{} \\
\textit{NormAI}                              & \multicolumn{2}{c}{} &  0.059          & 0.159 \\
\textit{UsedAI}{\(\times\)}\textit{NormAI}   & \multicolumn{2}{c}{} & -0.516\sym{**} & 0.197 \\
Sex (Male)                                   & -0.162\sym{*}   & 0.096 & -0.171\sym{*}   & 0.096 \\
Age                                          &  0.003          & 0.004 &  0.004          & 0.004 \\
University Education                         & -0.233\sym{**}  & 0.106 & -0.226\sym{**}  & 0.107 \\
Constant                                     &  3.115\sym{***} & 0.236 &  3.250\sym{***} & 0.233 \\
\midrule
Observations                                 & \multicolumn{2}{c}{245}   & \multicolumn{2}{c}{245} \\
$R^{2}$                                      & \multicolumn{2}{c}{0.290} & \multicolumn{2}{c}{0.294} \\
Root MSE                                     & \multicolumn{2}{c}{0.765} & \multicolumn{2}{c}{0.762} \\
F-statistic                                  & \multicolumn{2}{c}{19.11} & \multicolumn{2}{c}{23.10} \\
\bottomrule
\end{tabular}

\begin{tablenotes}\footnotesize
\item Robust standard errors in parentheses. \\ $^{*}p<.10$, $^{**}p<.05$, $^{***}p<.01$ after Holm-Benferroni correction.
\end{tablenotes}
\end{threeparttable}
\end{table}

The first regression model is used to test hypotheses H1, H2, H3, and H5. The second model is used to test H4 through the interaction term. Both regressions employ robust standard errors. Visual inspection of the standardized residual histograms and Q–Q plots revealed an approximately symmetric distribution. Both joint skewness-curtosis tests were non-significant at the 5\% level ($\chi^2 = 5.92, p = 0.052, \chi^2 = 3.16, p = 0.206$). All variance-inflation factors (VIFs) of the first model were $\approx 1.0$, but increased in the second model ($\text{Mean VIF} = 1.76$). Table \ref{tab:regression_models} presents the results of the two regression models. The p-values comparisons of the hypothesis variables were adjusted using the Holm–Bonferroni correction \citep{Holm_1979}.

In the first hypothesis, we expect regret to be higher when behavior deviates from the norm compared to when it adheres to the norm. The first model confirms this expectation ($\beta_{\text{CounterNorm}} = 0.25$, $t = 2.54$, $p = 0.012$). We also find a significant effect of AI usage on regret, which supports our second hypothesis that, in line with the concept of algorithm aversion, regret is higher when AI is used ($\beta_{\text{UsedAI}} = 0.89$, $t = 9.22$, $p = 0.000$).

According to the third hypothesis, we expected these two effects to differ---specifically, that the effect of counter-normative behavior would be stronger than that of AI usage. However, the significantly different coefficients $\beta_{\text{CounterNorm}} = 0.25$ and $\beta_{\text{UsedAI}} = 0.89$ suggest the opposite, which is also confirmed by a linear hypothesis test ($F(1, 238) = 18.48$, $p = 0.000$).

The fourth hypothesis addresses the interaction between norms and AI usage. We expected that regret from using AI would be reduced when AI usage aligns with the norm. The significant interaction term in the second regression model supports this hypothesis ($\beta_{\text{UsedAI} \times \text{NormAI}} = -0.52$, $t = -2.63$, $p = 0.009$).

The fifth and final hypothesis concerns the source of the norm. We hypothesized that when the perceived norm originates from a superior---as portrayed in the vignette through a managerial figure---it would lead to higher regret than when the norm is perceived through observation of colleagues. The first model shows no significant relationship between the source of the norm and regret ($\beta_{\text{SourceSuperior}} = 0.16$, $t = 1.62$, $p = 0.106$).

Regarding the control variables, we observe only a weakly significant relationship between \textit{regret} and sex or age. However, there is a significant negative relationship with university education: individuals with a university-level education report lower levels of regret ($\beta_{\text{UniversityEducation}} = -0.23$, $t = -2.11$, $p = 0.036$).

\subsection{Qualitative Analysis of Reasons and Justifications}
In the qualitative part of the experimental design, participants were asked in open-ended questions what they believed motivated the decision and how they would justify the choice if they were in the position of the new employee.

\begin{figure}[h!]
 \centering
        \includegraphics [width=1\textwidth]{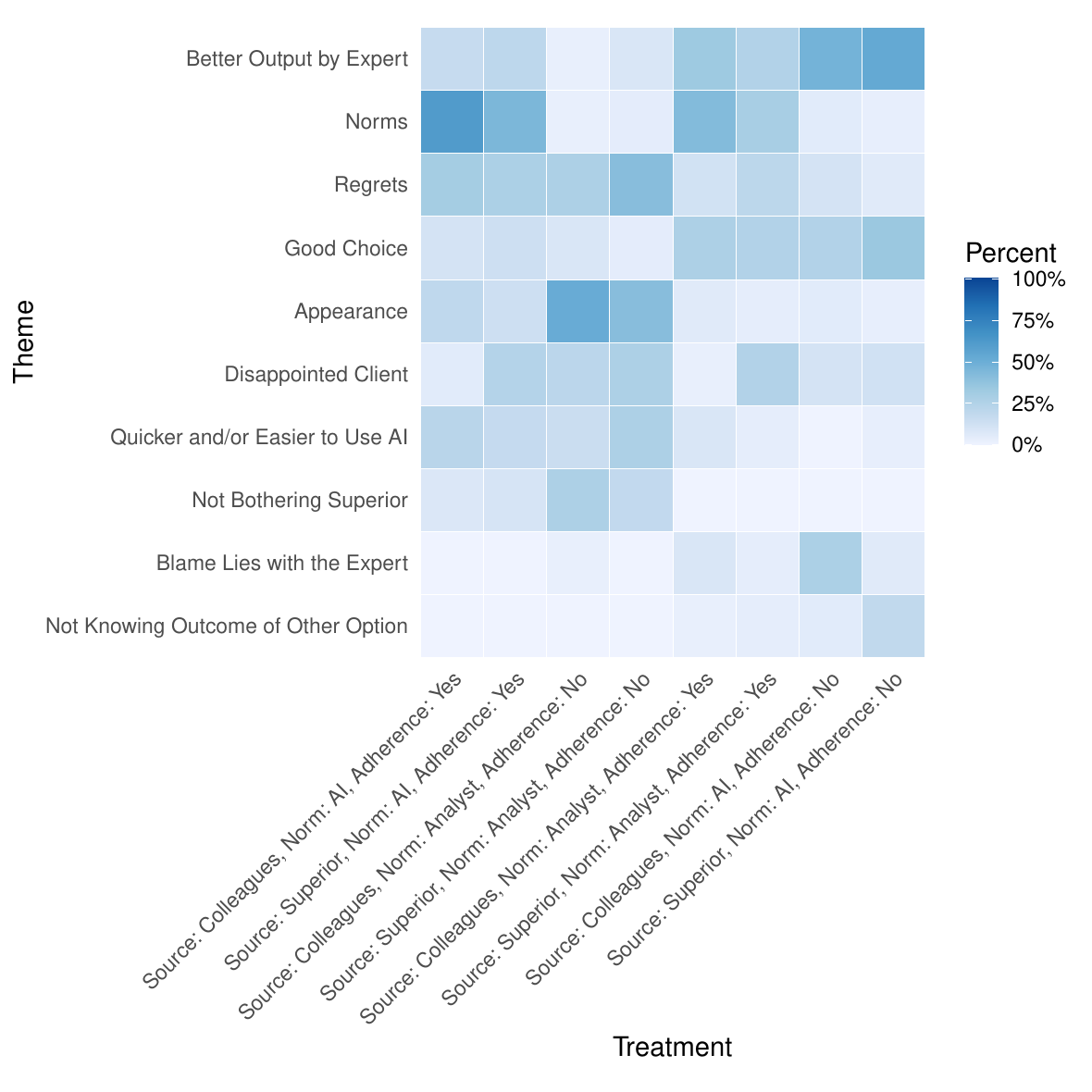}
        \caption{Themes regarding the question why participants regret their decision.}
        \label{fig:verbal}
        \end{figure}

Figure \ref{fig:verbal} illustrates the most frequently mentioned themes that respondents identified as justifications for the observed choice behavior. A notable pattern emerges: quality-related considerations (``Better output by experts'', ``Good choice'') appear more frequently when participants opted for the expert rather than the AI, both in cases of norm confirmation and norm deviation (last four treatments). Moreover, we observe that blame-shifting (``Blame lies with the expert'') emerges as a recurring theme in one specific treatment, namely when the prevailing norm favored AI but the participant chose the expert.

In contrast, expressions of regret are more prevalent when the AI was chosen over the expert—again in both norm-confirming and norm-deviating contexts (first four treatments). In these treatments, convenience-based justifications (``AI is quicker and easier'') as well as arguments of social convenience (“not bothering experts”) are mentioned more frequently.

        \begin{figure}[h!]
 \centering
        \includegraphics[width=1\textwidth]{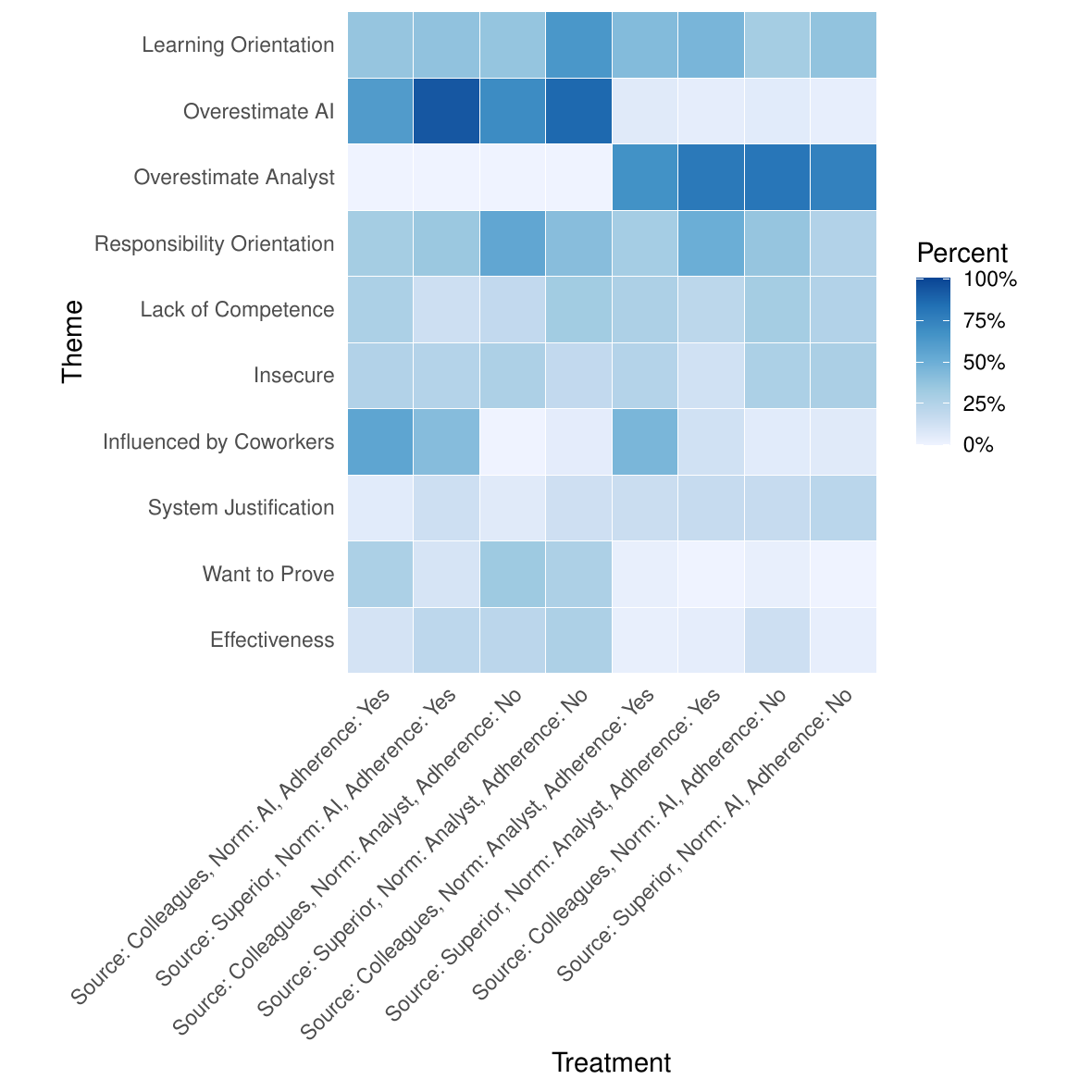}
        \caption{Themes regarding the justification of their choice.}
        \label{fig:justification}
        \end{figure}

Figure \ref{fig:justification} presents the most frequently mentioned themes when participants were asked how they would justify their behavior if they were in the position of the new employee. A striking pattern is the recurrence of blame-shifting strategies: individuals tended to attribute their choices either to an overestimation of the expert’s competence (``Overestimate authority'') or, conversely, to an overestimation of the AI’s abilities.

The second most prominent category relates to self-reported shortcomings, which were articulated in both positive and negative terms—ranging from “learning orientation” to ``lack of competence'' or ``insecurity.'' These types of justifications appear with roughly equal frequency across all treatments.

\section{Discussion}
AI is increasingly deployed as a decision-support system in an wide range of domains \citep{Rajpurkar_Chen_Banerjee_Topol_2022,Cao_2022,Dell_2023}. Yet, perceptions of and reliance on AI-generated recommendations vary widely  \citep{Mahmud_Islam_Mitra_2023}. A large strand of the literature seeks to reduce algorithmic aversion through interventions such as increased explanability \citep{Zhang_Liao_Bellamy_2020} or better tailoring to users \citep{Kawaguchi_2021}. However, attitudes towards and use of AI does not only depend on the technical system, but also on the social context. Our paper analyzes social norms that favor using AI for decision support as a potential intervention. Prior work has shown that AI-favoring norms can increase acceptance, but it has not addressed how such a norm might be established \citep{alexander2018trust,Bogard_Shu_2022}.

Using a vignette experiment, we examine which origin of an AI-favoring norm is more effective. We present a scenario in which a new employee has chosen between human advice and AI advice and perceives either their colleagues or their supervisor as the source of the norm. To assess which source is more influential, participants---acting in the role of the employee---reported their self-reported regret, after discovering that their decision led to an regrettable outcome.

First, counter-normative behavior (e.g., using AI instead of asking a human, and vice versa) is related to increased regret. This establishes a necessary precondition, demonstrating that regret is a suitable operationalization for testing the remaining hypotheses. This replicates the experimental findings of \cite{Bogard_Shu_2022}, who likewise use regret as the dependent variable. However, the effect size in their experiments is substantially larger, even after rescaling their measures to match ours. One possible reason is differences in item wording: whereas we used the \cite{Brehaut_2003} regret scale, \cite{Bogard_Shu_2022} relied on a single item measure in some experiments. Another possibility is that the decision domains differ: in medical and incentivized sports contexts, risk is higher, so regret following poor outcomes can be expected to be greater.

Second, we find that the choice of AI is related to increased  regret, confirming our hypotheses regarding algorithm aversion. This pattern aligns with \cite{Bogard_Shu_2022}, who also detect aversion, but the effect sizes differ. In our experiment, using AI has roughly triple the impact on regret compared with counter-normative behavior; in studies by \cite{Bogard_Shu_2022}, the relationship is reversed, with counter-normative behavior exerting the stronger effect. This difference shapes how norms interact with AI use. \cite{Bogard_Shu_2022} report a ``reversal'' of algorithm aversion: when using AI was normative and participants complied, they exhibited ``algorithm appreciation.'' We do not observe a reversal; rather, aversion is attenuated.

Although the results do not fully coincide, we agree that much of algorithm aversion has social roots. Consistent with prior work, acceptance of algorithms rises with broad adoption \citep{alexander2018trust}, and similar dynamics appear for other technologies \citep{huttel2022importance}. Thus, social norms can serve as an intervention to increase acceptance of AI-based decision-support systems when uptake falls short of the optimum. Averse users may be nudged via norm-framed descriptions; this can be effective even at low adoption levels, as shown in research about sustainable consumption \citep{Demarque2015nudgingnorms}. However, additional interventions are likely needed to further reduce potentially irrational aversion. These might include improving system transparency, providing explanations for AI-generated recommendations, or involving users more directly and tailoring the systems to their needs.

Third, for social norms to be used effectively as an intervention, they need a source that can generate a sufficiently high adoption rate to establish AI use as a norm. In hierarchical organizational settings, this source could be a superior uses AI, but it could also be fellow employees from whom the norm is observed. For those aiming to establish such a norm, the question is which source is more effective.

According to our results, the origin of the norm—whether it comes from a superior or from colleagues---does not have a significant impact. One possible reason for this null effect might be that the two sources may have appeared too similar within the vignette. It is possible that a superior would have a stronger effect if their higher status had been depicted more clearly, making them more salient social referents \citep{paluck2012salience} and strengthening the perception of the norm. Their elevated status might also support the establishment of an injunctive norm. While descriptive norms tend to be more influential \citep{Melnyk_van}, the combination of both types could prevent a null effect from the superior source \citep{Bhanot_2021}. However, in our experiment, we deliberately excluded injunctive norms to more clearly isolate the influence of descriptive norms from each source.

One factor that could potentially strengthen the effect of norms coming from colleagues is perceived similarity. Colleagues may resemble the employee more closely, and this sense of shared identity with the norm reference group can be influential \citep{Tankard_Paluck_2016,Goode_Balzarini_Smith_2014,Stangor_Sechrist_Jost_2001}. For instance, if some of the employees in the vignette were also relatively new to the organization, they might serve as particularly relevant norm referents. Their ``low-ranking'' position could additionally have a positive effect \citep{Dannals_Reit_Miller_2020}. Future studies could build on our experimental framework to explore these potential moderating factors and further clarify the difference between superior- and employee-driven norms. We assume that especially field studies should be used to investiagte the effects of hierarchy and social norms. 

A limitation of our study is that we did not analyze participant characteristics in great detail. Among our control variables, we found no association between age and regret, even though meta-analysis by \cite{Melnyk2022} indicates that older individuals are less susceptible to norms. Regarding gender, we found a small and weakly significant association, which aligns with the inconsistent findings in the existing literature \citep{Melnyk2022}. Other aspects that could also be examined include psychological characteristics that may have moderating effects, as certain dimensions of the Big Five personality traits, for example, are associated with norm adherence \citep{Zhao_Ferguson_Smillie_2017,Han_2021,Eck_Gebauer_2022}.

Our findings underscore the need for future research on AI diffusion, aversion, and acceptance to more systematically account for the role of the social context. The adoption of AI is not determined solely by the technical characteristics of the system or the dispositions of the individual user. Rather, it is also embedded in a broader organizational and societal environment, in which the prevailing attitudes of supervisors, colleagues, and other relevant referent groups can significantly shape patterns of use. This highlights the importance of understanding AI adoption as a socially situated process, rather than a purely individual---level or technology---driven phenomenon.

The qualitative evidence further draws attention to blame-shifting as a central theme in decision-making. Our results suggest that blame-shifting \citep{lozano2019effect} may operate as a psychological mechanism for mitigating regret, enabling individuals to deflect responsibility for undesirable outcomes. Interestingly, the salience of blame and blame-shifting appears to be contingent on the presence of social hierarchies. When social status matters, blame attribution becomes a particularly salient means of preserving self-image and mitigating reputational costs, which reinforces the broader point that AI-related choices are shaped by the social environment rather than existing in isolation \citep{malle2022cognitive}. This aligns with the idea that decision-makers may prefer to delegate to algorithms to avoid being blamed \citep{Maasland_Weißmuller_2022} and that human advisors are sometimes preferred over algorithms precisely because they can absorb blame \citep{onkal_Goodwin_Thomson_Gönül_Pollock_2009}.

Building on these insights, future research should investigate in greater detail how the availability of blame targets and the dynamics of blame-shifting influence both AI aversion and AI appreciation. One promising avenue is to examine whether making responsibility more explicit and salient can help foster more rational and deliberate decision-making regarding AI use. This is particularly relevant in organizational contexts, where ambiguity about responsibility may exacerbate algorithm aversion and reduce uptake. At the same time, personal characteristics—--such as personality traits, locus of control, or risk preferences--—may moderate these dynamics, shaping whether individuals are more inclined to deflect responsibility or to accept AI-generated recommendations.

Taken together, these considerations suggest that advancing our understanding of AI acceptance requires not only a focus on technical improvements (e.g., transparency or explainability) but also a deeper analysis of the social mechanisms that underpin decision-making. Integrating insights from social psychology, organizational behavior, and human–AI interaction research will be crucial to develop a more comprehensive framework of how norms, responsibility structures, and blame dynamics jointly shape the trajectory of AI diffusion.

\section{Conclusion}

Our study suggests that algorithm aversion is not merely a matter of system design or individual predisposition but is deeply embedded in social contexts. We show that counter-normative behavior amplifies regret and that AI use, despite being increasingly common, continues to evoke stronger regret than reliance on human advice. While the source of AI-favoring norms---whether colleagues or superiors---did not significantly alter this dynamic in our experiment, our findings suggest that social referents, blame-shifting, and perceptions of responsibility critically shape how AI recommendations are evaluated.

These results highlight the importance of understanding AI adoption as a socially situated process. Interventions that harness social norms, clarify responsibility, and address blame dynamics may complement technical solutions such as explainability and transparency. Future research should further explore how organizational hierarchies, group identities, and individual traits interact with these mechanisms to foster a more informed, deliberate, and responsible AI use.

%\section*{Acknowledgements}
%The author(s) would like to thank  

\section*{Data availability statement}
\label{sec:data}
The code for the program software, the experiment data, and the analysis code can be found in the public repository:\\\url{https://osf.io/u6wkg/?view_only=87666c46dc614e71904339968905fc20}.

\section*{Institutional review board statement} 
The ethics board of the first authors University has approved the research project.

\ifanonymous\else
\section*{Statements and Declarations}
The authors have nothing to declare.
\section*{Funding}
The authors gratefully acknowledges funding by the German Research Foundation (Deutsche Forschungsgemeinschaft, DFG): TRR 318/1 2021 – 438445824
\fi

% \end{multicols}
\bibliography{references}

\newpage
\appendix
\pagenumbering{roman} % switches to Roman numerals
\section{Appendix}

\subsection{Appendix - Instructions}
\label{appendix:instructions}

\textbf{Instructions}

Dear Participant,

Thank you for your interest in our study. This page will provide you with a detailed set of instructions to guide you through our study. Please read this carefully before starting.

\textbf{Study Overview}

On the following page, you will be presented with a scenario involving a new employee at a consulting firm. Your task is to put yourself in the employee’s position. To do this, please read through the scenario carefully and attentively.

It is important that you read through the scenario carefully in order to better put yourself in the shoes of the new employee. Once you have finished reading, you can click ``Next'' and then you will be asked to answer a few questions.

We appreciate your participation and commitment to this study. If you have any questions, please feel free to contact us.

\textbf{Comprehension Check}

To ensure that you have thoroughly understood these instructions, you will need to answer a set of comprehension questions. Please be aware that if you fail to answer one out of these questions correctly after three attempts, you will be unable to continue with the study.

\newpage
\subsection{Appendix - Screenshots}
\label{appendix:screenshots}

\begin{figure}[h!]
    \caption{Screenshot of the vignette for the treatment combination: Norm Adherence: Yes; Norm: Senior Analyst; Source of Norm: Colleagues}
    \label{fig:placeholder}
    \centering
    \includegraphics[width=0.7\linewidth]{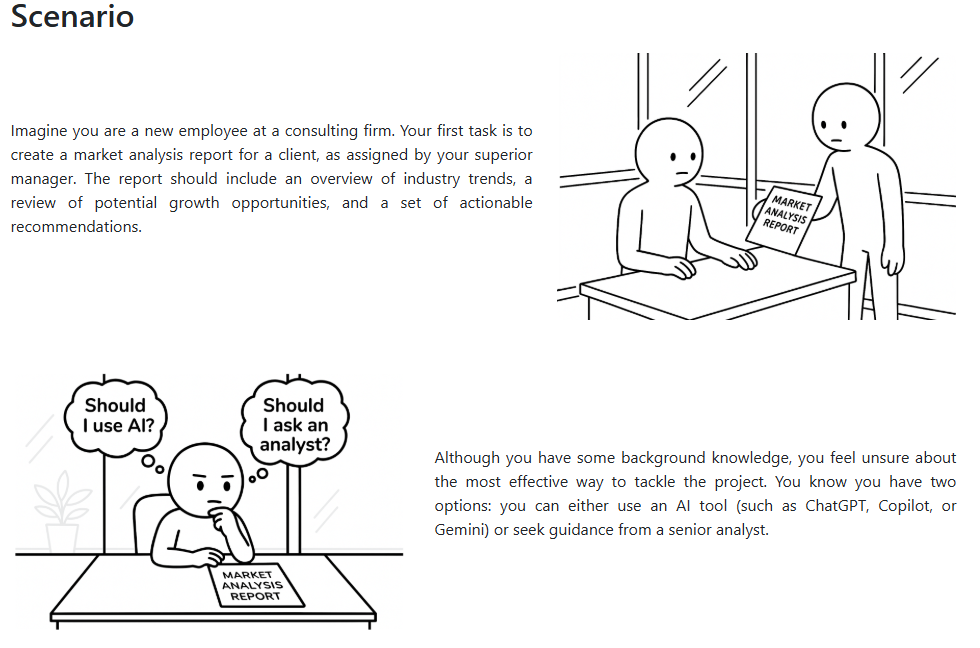}
\end{figure}

\begin{figure}[h!]
    \centering
    \includegraphics[width=0.7\linewidth]{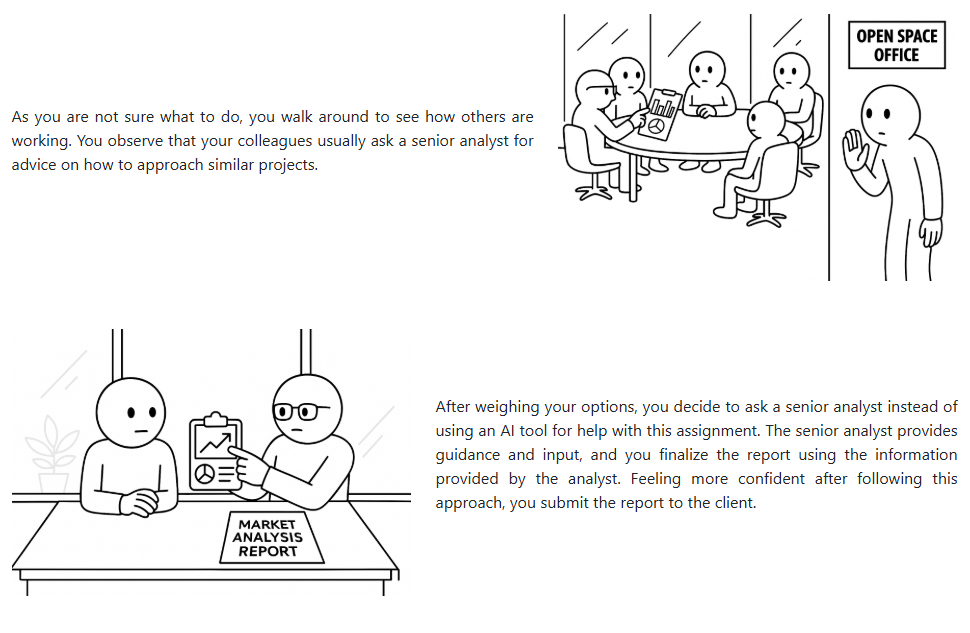}
    % \caption{Enter Caption}
    \label{fig:placeholder}
\end{figure}

\begin{figure}[h!]
    \centering
    \includegraphics[width=0.7\linewidth]{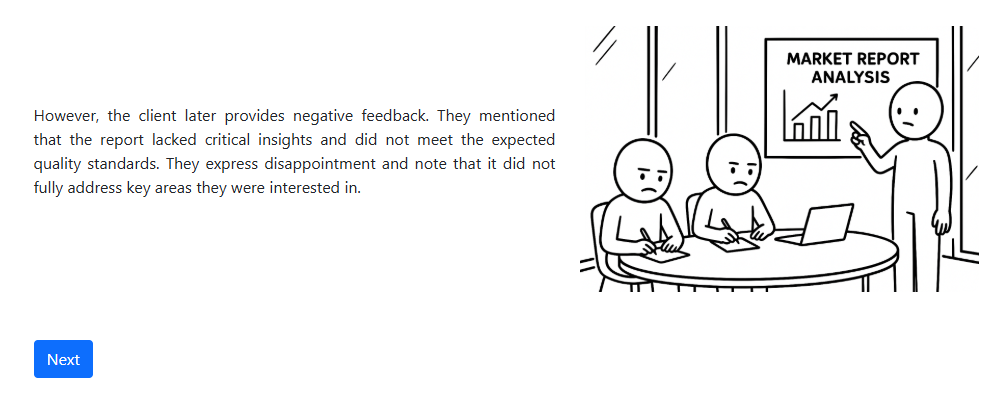}
    % \caption{Enter Caption}
    \label{fig:placeholder}
\end{figure}

\subsection{Appendix - Pairwise significance}
\label{appendix:pairwise}
\begin{landscape}
\begin{table}[p]\centering
\large
\setlength{\tabcolsep}{8pt}
\renewcommand{\arraystretch}{1.15}
\newcommand{\NA}{---}
\caption{Pairwise comparisons t-tests (lower triangle; blanks above the diagonal).}
\label{tab:pairwise}
\resizebox{1.5\textwidth}{!}{%
\begin{tabular}{l*{8}{c}}
\toprule
& \makecell[l]{Norm Adherence: No\\ Norm: AI\\ Source: Colleagues}
& \makecell[l]{Norm Adherence: No\\ Norm: AI\\ Source: Superior}
& \makecell[l]{Norm Adherence: No\\ Norm: Analyst\\ Source: Colleagues}
& \makecell[l]{Norm Adherence: No\\ Norm: Analyst\\ Source: Superior}
& \makecell[l]{Norm Adherence: Yes\\ Norm: AI\\ Source: Colleagues}
& \makecell[l]{Norm Adherence: Yes\\ Norm: AI\\ Source: Superior}
& \makecell[l]{Norm Adherence: Yes\\ Norm: Analyst\\ Source: Colleagues}
& \makecell[l]{Norm Adherence: Yes\\ Norm: Analyst\\ Source: Superior} \\
\midrule
\makecell[l]{Norm Adherence: No\\ Norm: AI\\ Source: Colleagues}
& \NA &  &  &  &  &  &  &  \\
\makecell[l]{Norm Adherence: No\\ Norm: AI\\ Source: Superior}
& 1.000 & \NA &  &  &  &  &  &  \\
\makecell[l]{Norm Adherence: No\\ Norm: Analyst\\ Source: Colleagues}
& .000 & .000 & \NA &  &  &  &  &  \\
\makecell[l]{Norm Adherence: No\\ Norm: Analyst\\ Source: Superior}
& .000 & .001 & 1.000 & \NA &  &  &  &  \\
\makecell[l]{Norm Adherence: Yes\\ Norm: AI\\ Source: Colleagues}
& .003 & .547 & .007 & .036 & \NA &  &  &  \\
\makecell[l]{Norm Adherence: Yes\\ Norm: AI\\ Source: Superior}
& .002 & .168 & .714 & 1.000 & 1.000 & \NA &  &  \\
\makecell[l]{Norm Adherence: Yes\\ Norm: Analyst\\ Source: Colleagues}
& 1.000 & 1.000 & .000 & .000 & .007 & .003 & \NA &  \\
\makecell[l]{Norm Adherence: Yes\\ Norm: Analyst\\ Source: Superior}
& 1.000 & 1.000 & .000 & .000 & .180 & .055 & 1 & \NA \\
\bottomrule
\end{tabular}%
}
\end{table}
\end{landscape}
\clearpage

\newpage
\subsection{Appendix - Coding Example}
\label{appendix:Coding}

\begin{table}[ht]
\centering
\caption{\textit{Example answers and frequency of the themes of the question about influencing factors.}}
\label{tab:example_answers_themes_q1}
\renewcommand{\arraystretch}{1.2}
\begin{tabular}{p{0.66\textwidth} p{0.24\textwidth} c}
\toprule
\textbf{Example Answer} & \textbf{Theme} & \textbf{Total} \\
\midrule
I chose a senior analyst because they should have many years of experience working for these clients and should be able to get the best inputs. & Better output by expert & 66 \\
I saw everyone else at the company using AI and, therefore, reasoned that it was standard practice and that it was seen as a tried, tested and trusted method. & Norms & 61 \\
I felt really bad, I should have known better and do better instead of getting negative report about my presentation. & Regrets & 52 \\
It was a reasonable choice. The negative feedback could well have been incorrect or insufficient prompting rather than anything wrong with AI in itself. & Good choice & 47 \\
I thought I might be negatively judged if my colleagues found out I had used AI. & Appearance & 43 \\
I was disappointed because my analysis did not meet the requirements from the requester. & Disappointed client & 37 \\
AI is quicker and in theory can provide deeper insights. But seriously, AI is actually pretty bad. & Quick and/or easier to use AI & 29 \\
Being new in a company and not feeling comfortable enough to ask superiors for assistance. & Not bothering superior & 19 \\
I feel that the senior analyst should have been able to give me good information and advice. & Blame lies with the expert & 17 \\
The fact I received negative feedback from the client. However, there is no way of knowing whether AI would've produced a better result. & Not knowing outcome of the other option & 10 \\
\bottomrule
\end{tabular}
\end{table}

\newpage

\begin{table}[ht]
\centering
\caption{\textit{Example answers and frequency of the themes of the question about justification.}}
\label{tab:example_answers_themes}
\renewcommand{\arraystretch}{1.2}
\begin{tabular}{p{0.66\textwidth} p{0.24\textwidth} c}
\toprule
\textbf{Example Answer} & \textbf{Theme} & \textbf{Total} \\
\hline
I soiught advise of a senior analyst as that is what i understand to be company practice. But I did not present it properly and seek coaching on this & Learning Orientation & 98 \\
I would explain that it seemed like the most effective way, but clearly lacked sufficient information to get optimum results & Overestimate AI & 97 \\
I would apologise and say I'm not sure why the analysis didn't go to plan, and I will ask them next time would it be better if I used AI? & Overestimate analyst & 91 \\
I would explain that I decided to use AI and it was probably the wrong decision to take & Responsibility orientation & 91 \\
I would explain that due to lack of experience I made a poor choice that I deeply regret and that I will learn from this. & Lack of cpetence & 60 \\
i regret that my report did not have all of the information needed. i was not 100\% sure on how to do the task and should have asked for help. & Insecure & 59 \\
I wasn't sure on what approach I should use and as I saw it being used by others I thought it would be ok but I should of asked first which I know now & Influenced by coworkers & 55 \\
I believe the analysis fell short because i lacked clear guidance or resources, leading to misunderstandings in meeting the client's expectations & System justification & 33 \\
i used my initaative as i was unsure on the task. having seen yourself aand colleuges use the ai tool i used this same tool to help orduce my work rather thn come to you with a problem & Want to Prove & 32 \\
I would try to explain that I thought that using the AI tool would improve my knowledge and give me the information needed so that I wouldn't look out of my depth to a senior analyst. & Effectiveness & 31 \\
\bottomrule
\end{tabular}
\end{table}

\end{document}